\documentclass[11pt,british]{scrartcl}
\usepackage[T1]{fontenc}
\usepackage[utf8]{inputenc}
\usepackage[english]{babel}
\usepackage{geometry}
\usepackage{units}
\usepackage{amsmath}
\usepackage{graphicx}
\usepackage{color}

\begin{document}
\title{Interacting bosons in two-dimensional flat band systems}

\author{Petra Pudleiner, \\
Institut für Physik, University of Mainz, Germany\\
Andreas Mielke, \\
Institut für Theoretische Physik, University of Heidelberg, Germany}
\maketitle
\begin{abstract}
The Hubbard model of bosons on two dimensional lattices with a lowest
flat band is discussed. In these systems there is a critical density,
where the ground state is known exactly and can be represented as
a charge density wave. Above this critical filling, depending on the
lattice structure and the interaction strength, the additional particles
are either delocalised and condensate in the ground state, or they
form pairs. Pairs occur at strong interactions, e.g., on the chequerboard
lattice. The general mechanism behind this phenomenon is discussed.
\end{abstract}

\section{Introduction}
\label{intro}
Flat band systems have been studied extensively in experiment and
theory. They are a prototype for strongly correlated systems. Strongly
correlated phases of matter emerge in such systems since the interaction,
even if it is small, dominates the behaviour. The interaction leads
to different phases. Fermions in flat bands are known to
show ferromagnetism \cite{Mielke1991,Tasaki92,Mielke1993} and ferrimagnetism
\cite{Lieb89}, for a review we refer to \cite{Tasaki97}.
More recently, bosons in flat bands have been studied, esp. the
question of Bose condensation \cite{Huber2010} and the 
emergence of topologically ordered phases such as lattice versions
of fractional quantum Hall states (see e.g. \cite{wang2011fractional,wang2012non,1206.3759v3})
are important here. Hard-core bosons can be mapped to spin systems, which
as well have been investigated, see e.g. \cite{Schulenburg2002,Zhitomirsky2004}. 

Promising experiments to realise such systems are ultra-cold atoms
in optical lattices; as they provide a perfect implementation of the
Hubbard model. Optical lattices might be viewed as quantum simulators
and are realised by coun{-}ter-propagating laser beams forming a periodic
microtrap for the atoms. These experiments enable the control of a
large number of parameters as the potential depth or the lattice geometry
itself, see e.g. \cite{greiner2002QuaphatrasuptoMotinsgasultato,bloch2005Ultquagasoptlat,lewenstein2007Ultatogasoptlatmimconmatphybey_2,bloch2008Manphyultgas}. 

For bosons in a flat band, different phases may occur. A especially
interesting question is whether Bose condensation occurs. This question
was, to our knowledge, first investigated by Huber et al. \cite{Huber2010}
for bosons on a kagom{\'e} lattice. The kagom{\'e} lattice is the line graph
of the hexagonal lattice \cite{Mielke1991} and thus a prototype of
a larger class of lattices. For these and similar lattices, the dimensionality
is important. This was already shown by Huber et al. \cite{Huber2010}.
They compared the one-dimensional saw-tooth chain, the line graph
of a decorated chain, with the kagom{\'e} lattice. Both lattices have
an ordered ground state at some critical filling $\rho_{c}$: a charge
density wave (CDW). Adding further particles to the CDW destroys the
order in the one-dimensional case. Domain walls appear, which can
freely move, and the system becomes a Luttinger liquid \cite{Huber2010}.
This result holds at weak interaction. In the hard-core limit, a two-body
bound state is formed which can move through the lattice \cite{Takayoshi2013,Tovmasyan2013,Phillips2014}.
On the kagom{\'e} lattice, in contrast, domain walls are less favourable
than extended states. The CDW ground state on the kagom{\'e} lattice is
three-fold degenerate. Each of these states has interstitial sites
which are unoccupied. The additional particles have a high mobility
and probability to move on the interstitial sites. A small
but finite density of additional particles form a condensate. The
underlying CDW states remain intact for densities above but close to
the critical density and break down at higher densities \cite{Huber2010}.
These results hold for weakly interacting bosons. For strongly interacting
bosons, esp. for hard-core bosons, other states may be favourable.

Several questions arise from these results. Since the kagom{\'e} lattice
is a prototype of a larger class of lattices, one may ask if
the results are true on other line graphs of two dimensional bipartite
lattices. Another important member of this class is the chequerboard
lattice. For the chequerboard lattice, the CDW state at the critical
density is two-fold degenerate and there are no interstitial sites
\cite{1112.0131v3}. The first question therefore arises, whether
on the chequerboard lattice the physics of weakly interacting bosons
is the same as on the kagom{\'e} lattice. A second question is whether
for strongly interacting bosons on one of these lattices bound states
occur, as in the one-dimensional system. A third question is, what
happens in three dimensions. 

In this paper, we concentrate on line graphs of two dimensional bipartite
lattices, for several reasons. On this class of lattices, the flat
band contains strictly localised states. We will briefly sketch them
and their properties in the next section. For details we refer to
\cite{1112.0131v3,Mielke1992a}. Each finite face of the 
two dimensional bipartite lattices is surrounded by a cycle with an
even number of sites. In \cite{Mielke1992a} it was
shown that the states which are localised on these cycles
are linearly independent and complete. They form a basis, but not an orthonormal
basis, which makes it difficult to deal with them. For that reason,
we follow the idea of Huber et al. \cite{Huber2010} and introduce
a Wannier basis. The Wannier states are as well localised, but not
strictly on few lattice sites. They fall of algebraically. Similar to the approach
by Huber et al. we use the Wannier states in Sect. \ref{sec:weak-ia}
to derive an effective Hamiltonian for weakly interacting bosons on
the chequerboard lattice. It turns out that the essential physics
for weak interactions is the same as on the kagom{\'e} lattice. 

Let us mention that the algebraic decay of the Wannier states
was as well used by Chalker et al. \cite{Chalker2010} in
their investigation of disorder and localisation in two-dimensional
flat-band systems.

In Sect. \ref{sec:strong-ia} we study the case of hard-core bosons.
We use different kinds of variational states, partly by using exact
diagonalisations of small separable sub-units of the lattice. On the
chequerboard lattice, it turns out that pairs of bosons are formed.
On the kagom{\'e} lattice, the pair states exist as well, but have
not the lowest energy. The ground state on small separable sub-units
of the lattice shows a larger expectation value of the particle number
on the interstitial sites. This indicates that on the kagom{\'e} lattice
the physical picture developed by Huber et al. remains true for strong
interactions. The physics of hard-core bosons on the two lattices,
kagom{\'e} and chequerboard, is thus different. Finally, in Sect. \ref{sec:Conclusion-and-outlook}
we give some conclusions and discuss open questions.

\section{The model}
\label{sec:model}
\begin{figure*}[hb]
\begin{centering}
\begin{tabular}{ccc}
\includegraphics[height=0.3\columnwidth]{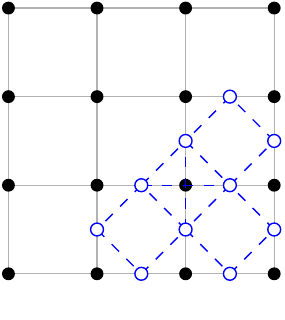} & \includegraphics[height=0.3\columnwidth]{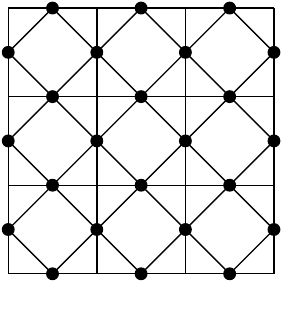} & \includegraphics[height=0.3\columnwidth]{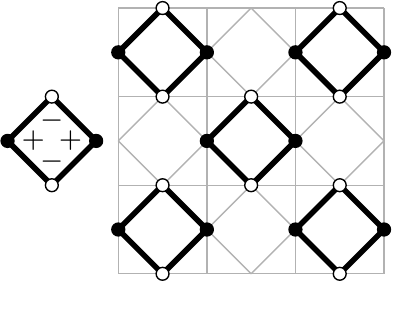}\tabularnewline
square lattice & chequerboard lattice & localised states forming a CDW state\tabularnewline
\end{tabular}
\par\end{centering}

\protect\caption{\label{fig:Illustration-of-the}Illustration of the construction of
the chequerboard lattice as the line graph of the square lattice and
localised states belonging to the faces. }

\end{figure*}

The Bose-Hubbard model on a lattice is defined by the Hamiltonan 
\begin{equation}
H=\sum_{x,y\in V}t^{\phantom{\dagger}}_{xy}b_{x}^{\dagger}b_{y}^{\phantom{\dagger}}+\frac{U}{2}\sum_{x\in V}b_{x}^{\dagger}b_{x}^{\dagger}b^{\phantom{\dagger}}_{x}b^{\phantom{\dagger}}_{x},\label{eq:bose_hubbard}
\end{equation}
where $V$ is the set of lattice sites. The interaction $U>0$ is
repulsive. We restrict ourselves to nearest neighbour hoppings, i.e.,
$t_{xy}=ta_{xy}$ where $A=(a_{xy})_{x,y\in V}$ is the adjacency
matrix of the lattice, i.e., $a_{xy}=1$ if $x,y$ are nearest neighbours,
$a_{xy}=0$ otherwise. We let $t>0$ in order to obtain a lowest flat
band in the systems we study. We restrict ourselves to lattices which
are line graphs of planar, bipartite lattices. Prominent examples
are the kagom{\'e} lattice, the line graph of the hexagonal lattice, or
the chequerboard lattice, the line graph of the square lattice. The
usual fermionic Hubbard model has been studied extensively on these
lattices and has ferromagnetic ground states \cite{Mielke1992a}.
The important feature of line graphs is that the spectrum of the
adjacency matrix is bounded from below by $-2$ and that the eigenvalue
$-2$ has a large degeneracy. Let $|V|$ be the number of lattice
sites and $n$ the number of lattice sites per unit cell, then
the degeneracy is $|V|/n+1$. $|V|/n$ is the number of states in
the lowest band. The additional state is an element of the next band,
which touches the lowest flat band. There is a one-to-one correspondence
between the faces of the original planar, bipartite lattice and the
localised eigenstates of $A$. Each face has, since the original lattice
is bipartite, an even number of edges. Each edge is, by construction,
a vertex of the line graph. A single particle state which has a constant
modulus on these edges and an alternating sign is an eigenstate of
$A$ with eigenvalue $-2$. The eigenstates are linearly independent
and thus form a basis of the eigenspace belonging to the eigenvalue $-2$.
Fig. \ref{fig:Illustration-of-the} illustrates the procedure for
the chequerboard lattice. 

Let us denote the original bipartite planar lattice as $G$. It consists
of two sub-lattices, $G_{1}$ and $G_{2}$. Formally, each edge of
the original bipartite planar lattice can be oriented to point from
$G_{1}$ to $G_{2}$. Further, each face can be oriented clockwise.
Since the vertices $x$ correspond to edges in $G$, we define for
each face $f$ 
\begin{equation}
s_{fx}=\begin{cases}
1, & \text{if }x\text{ belongs to the}\\
 & \text{boundary of }f\text{ and points}\\
 & \text{into the direction of }f,\\
-1, & \text{if }x\text{ belongs to the}\\
 & \text{boundary of }f\text{ and points}\\
 & \text{into the opposite direction}\\
 & \text{of }f,\\
0, & \text{otherwise,}
\end{cases}
\end{equation}
and the creation operators 
\begin{equation}
b_{f}^{\dagger}=\frac{1}{\sqrt{|f|}}\sum_{x\in V}s_{fx}b_{x}^{\dagger}.\label{eq:face_op}
\end{equation}
Due to the local structure of $b_{f}^{\dagger}$, it is possible to
construct multi-particle ground states for the Bose-Hubbard Hamiltonian
for densities below some critical density $\rho_{c}$. For low densities,
there exist many such states. Each set $F$ of non-touch{-}ing faces yields
a ground state $|F\rangle=\prod_{f\in F}b_{f}^{\dagger}|0\rangle.$
In fact, one can show that these states only form a subset of all
ground states at low densities. A complete rigorous description of
all multi-particle ground states below a certain critical density
$\rho_{c}$ has been given in \cite{1112.0131v3} for all line graphs
of planar bipartite lattices. The critical density is given by a close
packing of non-touching faces. Depending on the lattice, there may
be more than one close packing. The chequerboard lattice has two such
close packings. This procedure yields thus one or more charge density
wave (CDW) states. On the chequerboard lattice, the critical density
is $\rho_{c}=\frac{1}{4}$. There are two such states, one of
them is depicted in Fig. \ref{fig:Illustration-of-the}. We denote
the corresponding two subsets by $F_{1}^{c}$ and $F_{2}^{c}$.

Above the critical density it has to be examined whether the CDW is
destroyed completely or is disturbed merely locally. Therefore, the
energies of these configurations have to be compared for a lattice
filling with one further particle. In the case of the kagom{\'e} lattice
and for weak interactions, Huber et al. \cite{Huber2010} showed that
the additional particles become delocalised and form a condensate.
The kagom{\'e} lattice has a critical density of $\rho_{c}=\frac{1}{9}$ and
three degenerate CDW ground states at the critical filling. As a consequence,
a CDW state on the kagom{\'e} lattice has interstitial sites, which
belong to no occupied face. In this aspect, it differs to the chequerboard
lattice, on which a CDW state has no interstitial sites. 

\section{Weakly interacting model}
\label{sec:weak-ia}
In this section, we study the Bose Hubbard model on the chequerboard
lattice for weak interaction using a variational approach similar
to the one used by Huber et al. \cite{Huber2010} for the kagom{\'e} lattice.
This section contains no essentially new results, we just confirm
that the results in \cite{Huber2010} can be extended to the chequerboard
lattice and presumably to any other lattice which is a line graph
of a planar bipartite lattice. 

\begin{figure}[bh]
\begin{centering}
\includegraphics[width=0.8\columnwidth]{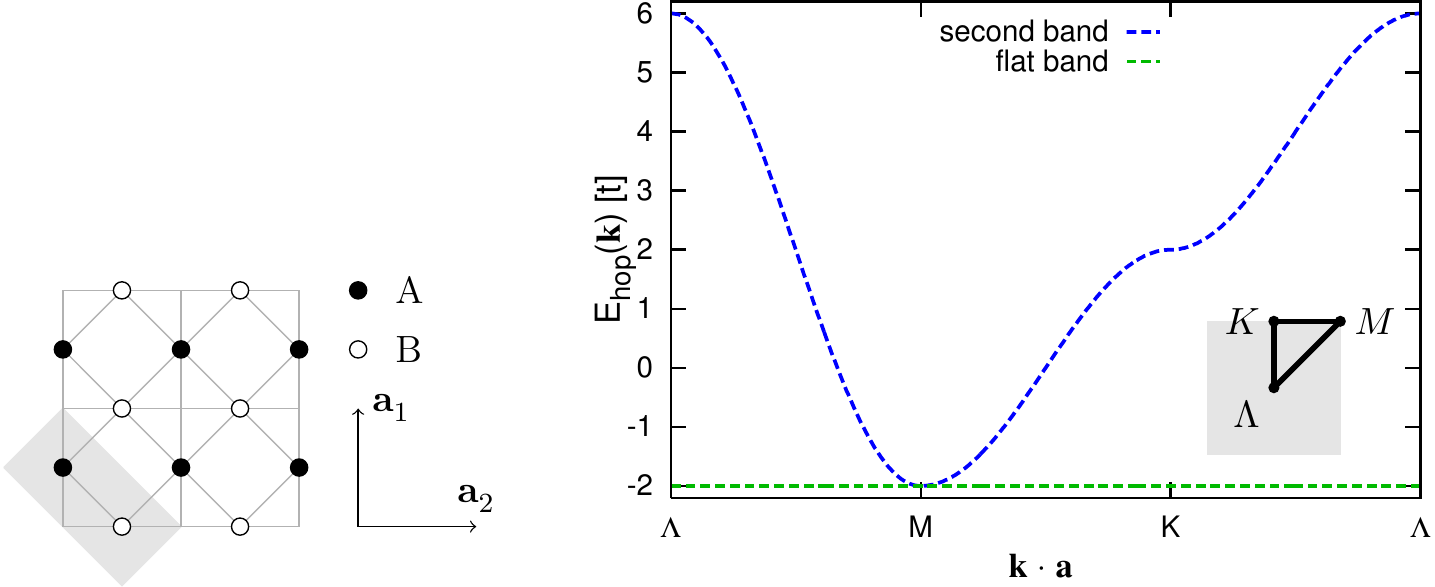}
\par\end{centering}

\protect\caption{\label{fig:The-band-structure}The band structure of the chequerboard
lattice in the first Brillouin zone (right) and the choice of the
unit cell (left) with the two lattice vectors $\mathbf{a}_{1,2}$ and the two lattice
types A and B.}
\end{figure}

The main problem with the basis states $b_{f}^{\dagger}$ constructed
above is that these states are not orthogonal. To keep the property
of localised states and to deal with an orthonormal basis, we follow
Huber et al. and construct a basis of Wannier states. This allows us
to project to the lowest band and to derive an effective Hamiltonian,
which can be treated further. As a starting point, we write the
hopping part of the Hamiltonian \eqref{eq:bose_hubbard} as
\begin{align}
H_{\text{hop}} & =t\sum_{\mathbf{k}}\left(b_{A,\mathbf{k}}^{\dagger},b_{B,\mathbf{k}}^{\dagger}\right)\nonumber\\
&\cdot\begin{pmatrix}z_{1}\left(k_{1}\right)-2 & z_{2}^{*}\left(k_{1},k_{2}\right)\\
z_{2}\left(k_{1},k_{2}\right) & z_{1}\left(k_{2}\right)-2
\end{pmatrix}\begin{pmatrix}b_{A,\mathbf{k}}\\
b_{B,\mathbf{k}}
\end{pmatrix}\:,\label{eq:diagHhop}
\end{align}
with 
\begin{align}
z_{1}\left(k_{\nu}\right) & =e^{ik_{\nu}}+e^{-ik_{\nu}}+2\\
z_{2}\left(k_{1},k_{2}\right) & =1+e^{ik_{1}}+e^{-ik_{2}}+e^{i\left(k_{1}-k_{2}\right)},
\end{align}
using $k_{\nu}=\mathbf{k}\cdot\mathbf{a}_{\nu}$ for $\nu\in\{1,2\}$. The vector $\mathbf{k}$ belongs to the first Brillouin zone. The
two vectors $\mathbf{a}_{\nu}$ are shown in Fig. \ref{fig:The-band-structure}.
The eigenvalues are $-2t$ and $2t\left[1+\cos\!\left(k_{1}\right)+\cos\!\left(k_{2}\right)\right]$
and are plotted in Fig. \ref{fig:The-band-structure} as well.

We now construct the Wannier basis for the lower flat band. The Wannier
operators are
\begin{eqnarray}
W_{i}^{\dagger} &=&\sum_{j}\Big[w_{A}^{*}(\mathbf{r}_{j}-\mathbf{r}_{i})b_{A,j}^{\dagger} \nonumber \\
 & & \quad +w_{B}^{*}(\mathbf{r}_{j}-\mathbf{r}_{i})b_{B,j}^{\dagger}\Big]\label{eq:localWann}
\end{eqnarray}
\begin{eqnarray}
w_{A}(\mathbf{r}_{i}) &=&  \int\frac{d^{2}k}{(2\pi)^{2}}e^{i\mathbf{k}\mathbf{r}_{i}} \nonumber \\
 & & \quad \frac{1+e^{-ik_{2}}}{\left[4+2\cos\!\big(k_{1}\big)+2\cos\!\big(k_{2}\big)\right]^{\frac{1}{2}}}\label{eq:localWanna}
\end{eqnarray}
\begin{eqnarray}
w_{B}(\mathbf{r}_{i}) &=& -\int\frac{d^{2}k}{(2\pi)^{2}}e^{i\mathbf{k}\mathbf{r}_{i}} \nonumber \\
 & & \quad \frac{1+e^{-ik_{1}}}{\left[4+2\cos\!\big(k_{1}\big)+2\cos\!\big(k_{2}\big)\right]^{\frac{1}{2}}},\label{eq:localWannb}
\end{eqnarray}
with the lattice vector $\mathbf{r}_{i}=\sum_{j}m_{j\nu}\mathbf{a}_{\nu}$
and integers $m_{i\nu}$. The index $i$ runs over all elementary
cells of the lattice. The creation operators $b_{A/B,i}^{\dagger}$
create bosons on the lattice site $A$ or respectively $B$ in the elementary cell
$i$. This state $W_{i}^{\dagger}$ involves bosons of all lattice
sites, weighted with the appropriate coefficients which depend on the
distance to the elementary cell $i$. These coefficients decay with
increasing distance at least by $\nicefrac{1}{\left|\mathbf{r}\right|}$
. 

Using the Wannier states, we write the local operators as 
\begin{align}
b_{A(B),i}^{\dagger}&=\sum_{j}\big[w_{A(B)}(\mathbf{r}_{i}-\mathbf{r}_{j})W_{j}^{\dagger}\nonumber\\
&\quad+\text{higher band}\big].\label{eq:inversWann}
\end{align}
Projecting onto the flat band means neglecting the contribution of
the higher band. This corresponds to the low-energy regime for a weakly
interacting system. We insert Eq. \eqref{eq:inversWann} into the Hamiltonian
\eqref{eq:bose_hubbard} and remove the energy offset $-2tN$. The
largest term in the resulting Hamiltonian is an on-site repulsion.
It is much larger than all the other terms. Following Huber et al.
\cite{Huber2010}, we replace it by a hard core interaction. The remaining
terms are interactions and assisted hoppings. Due to the decay of
the Wannier functions, the coefficients decay with increasing distance.
The first leading terms are
\begin{eqnarray}
H_{\text{eff}}^{(\text{w})} &=& P_{0}U\sum_{i,\nu\in(1,2)}W_{i}^{\dagger}W_{i}\bigg\{ \nonumber \\
 & & A_{1,i} + [ A_{2,i} +\text{h.c.}] \bigg\} P_{0},\label{eq:HeffSmallU}
\end{eqnarray}
where
\begin{eqnarray}
A_{1,i} &=& \frac{I_{1}}{2}W_{i\pm\mathbf{a}_{\nu}}^{\dagger}W_{i\pm\mathbf{a}_{\nu}}+2I_{2}W_{i\pm\mathbf{a}_{\nu}}^{\dagger}W_{i\mp\mathbf{a}_{\nu}}\nonumber \\
 &+& 2I_{3}W_{i\pm\mathbf{a}_{\nu}}^{\dagger}\Big[W_{i\pm\mathbf{a}_{\nu+1}}+W_{i\mp\mathbf{a}_{\nu+1}}\Big] 
\end{eqnarray}
\begin{eqnarray}
A_{2,i} &=& I_{2}W_{i\pm\mathbf{a}_{\nu}}^{\dagger}W_{i\pm2\mathbf{a}_{\nu}}\nonumber \\
 &+& I_{3}W_{i\pm\mathbf{a}_{\nu}}^{\dagger}\Big[W_{i\pm\mathbf{a}_{\nu}\pm\mathbf{a}_{\nu+1}} +W_{i\pm\mathbf{a}_{\nu}\mp\mathbf{a}_{\nu+1}}\Big]\nonumber \\
 &+& I_{4}W_{i\pm\mathbf{a}_{\nu}}^{\dagger}\Big[W_{i\mp\mathbf{a}_{\nu}\mp\mathbf{a}_{\nu+1}} +W_{i\mp\mathbf{a}_{\nu}\pm\mathbf{a}_{\nu+1}}\nonumber \\
 & & +W_{i\pm2\mathbf{a}_{\nu}\pm\mathbf{a}_{\nu+1}} +W_{i\pm2\mathbf{a}_{\nu}\mp\mathbf{a}_{\nu+1}}\Big]\nonumber \\
 &+& I_{5}W_{i\pm\mathbf{a}_{\nu}}^{\dagger}\Big[W_{i\pm3\mathbf{a}_{\nu}} +W_{i\mp2\mathbf{a}_{\nu}}\Big]
\end{eqnarray}

The coefficients are
\begin{alignat}{1}
\begin{aligned}
I_{1} & =2w_{A}(0)^{4} &  & \approx0.105\\
I_{2} & =2w_{A}(0)^{3}w_{A}(-\mathbf{a}_{2}) &  & \approx-0.018\\
I_{3} & =2w_{A}(0)^{3}w_{A}(\mathbf{a}_{1}) &  & \approx-0.013\\
I_{4} & =2w_{A}(0)^{3}w_{A}(\mathbf{a}_{1}-\mathbf{a}_{2}) &  & \approx0.008\\
I_{5} & =2w_{A}(0)^{3}w_{A}(-2\mathbf{a}_{2}) &  & \approx0.006\:.
\end{aligned}
\label{eq:abcoeff}
\end{alignat}
$P_{0}$ projects onto states with not more than one boson in a state
$W_{i}^{\dagger}$. 

Again following Huber et al. \cite{Huber2010}, we use a variational
ansatz for the ground state above $\rho_{c}$ of the form
\begin{align}
|\vartheta_{1}\vartheta_{2}\varphi_{1}\varphi_{2}\rangle&=\prod_{\alpha=1}^{2}\prod_{i\in F_{\alpha}}\Big[\cos(\vartheta_{\alpha}/2)\nonumber\\
&\quad+e^{i\varphi_{\alpha}}\sin(\vartheta_{\alpha}/2)W_{i}^{\dagger}\Big]|0\rangle.
\end{align}
The ansatz contains two CDW states $|\pi000\rangle=W_{1}^{\dagger}|0\rangle$
and $|0\pi00\rangle=W_{2}^{\dagger}|0\rangle$ with $W_{\alpha}^{\dagger}=\prod_{i\in F_{\alpha}}W_{i}^{\dagger}$,
both with filling $\rho_{c}$. Taking the expectation value of the
Hamiltonian \eqref{eq:HeffSmallU} yields 
\begin{align}
&\langle\Psi|H_{\text{eff}}^{(\text{w})}-\mu\hat{N}|\Psi\rangle =U\bigg[2A_{1}\sin^{2}\frac{\vartheta_{1}}{2}\sin^{2}\frac{\vartheta_{2}}{2}\nonumber\\
&+B_{1}\left[\sin^{2}\vartheta_{1}\sin^{2}\frac{\vartheta_{2}}{2}+\sin^{2}\vartheta_{2}\sin^{2}\frac{\vartheta_{1}}{2}\right]\nonumber \\
 &+B_{2}\left[\sin^{2}\vartheta_{1}\sin^{2}\frac{\vartheta_{1}}{2}+\sin^{2}\vartheta_{2}\sin^{2}\frac{\vartheta_{2}}{2}\right]\nonumber \\
 &+B_{3}\cos(\varphi_{1}-\varphi_{2})\sin\vartheta_{1}\sin\vartheta_{2}\left[\sin^{2}\frac{\vartheta_{1}}{2}\right.\nonumber\\
&\left. +\sin^{2}\frac{\vartheta_{2}}{2}\right]\bigg]\mbox{} -\frac{\mu}{2}\left[\sin^{2}\frac{\vartheta_{1}}{2}+\sin^{2}\frac{\vartheta_{2}}{2}\right].\label{eq:exp-order-0-01}
\end{align}
The coefficients are combinations of the interaction coefficients
$I_{i}$. They have the following values
\begin{alignat}{1}
\begin{aligned}A_{1} & \approx0.0527 & \quad\quad\quad B_{2} & \approx-0.0002\\
B_{1} & \approx-0.0138 & \quad\quad\quad B_{3} & \approx-0.0072.
\end{aligned}
\label{eq:coef-order0-01}
\end{alignat}

Eq. (\ref{eq:exp-order-0-01}) is minimised numerically by varying the
parameters $\vartheta_{1,2}$, $\varphi_{1}-\varphi_{2}$ and $\mu$; 
the latter in order to keep the particle number fixed. The density $\rho$
is varied between the critical filling factor $\rho_{c}=\frac{1}{4}$
and $\rho=\frac{1}{2}$. Starting point is the CDW state $W_{1}^{\dagger}|0\rangle$
at $\rho_{c}$. This enables to determine quantitatively the expectation
value of $H_{\text{eff}}^{(\text{w})}$ and by analogy to \cite{Huber2010}
the order parameter of the CDW
\begin{align}
\psi_{\text{CDW}} &=\frac{\langle\Psi|W_{1}^{\dagger}W_{1}-W_{2}^{\dagger}W_{2}|\Psi\rangle}{\langle\Psi|W_{1}^{\dagger}W_{1}+W_{2}^{\dagger}W_{2}|\Psi\rangle}\nonumber\\
&=\frac{\cos\vartheta_{1}-\cos\vartheta_{2}}{\cos\vartheta_{1}+\cos\vartheta_{2}-2},\label{eq:ordcdw}
\end{align}
and of the superfluid
\begin{align}
\psi_{\text{sf}} & =\frac{2}{N}\Big|\langle\Psi|W_{2}^{\dagger}|\Psi\rangle\Big|=\frac{1}{2}\Big|\sin\vartheta_{2}\Big|.\label{eq:ordsf}
\end{align}
\begin{figure}[hb]
\begin{centering}
\includegraphics[width=0.5\columnwidth]{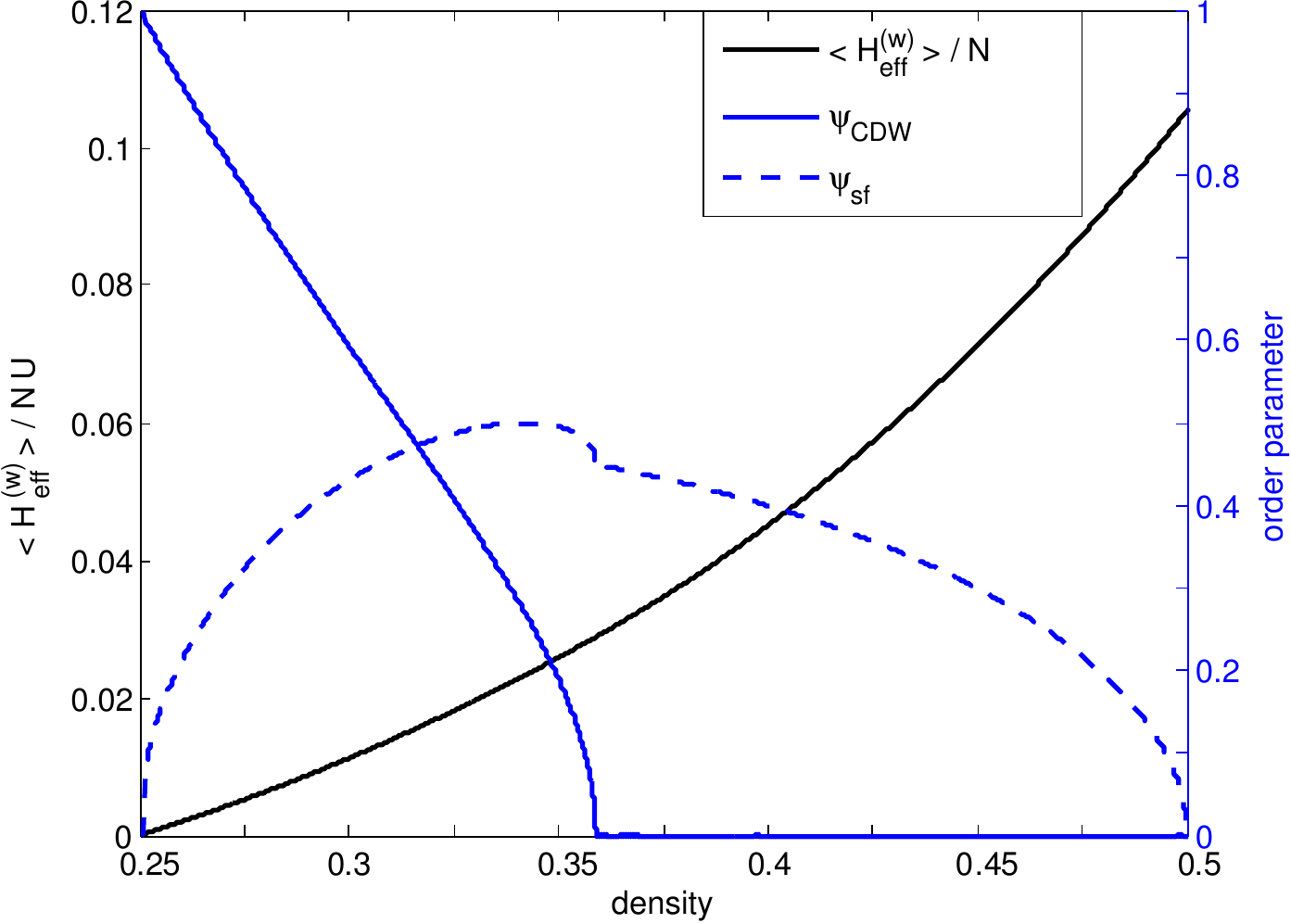}
\par\end{centering}

\protect\caption{\label{fig:mean-field-weak-ia}The expectation value of $H_{\text{eff}}^{(w)}$
[cf. Eq. \eqref{eq:exp-order-0-01}] is plotted on the left y-axis for the parameters obtained by solving the variational
problem. Eq. \eqref{eq:ordcdw} and \eqref{eq:ordsf} are likewise depicted
on the right y-axis for the order parameter of the CDW and the superfluid,
respectively.}

\end{figure}

The results are plotted as a function of the density in Fig. \ref{fig:mean-field-weak-ia}.
The order parameter of the CDW decreases continuously from $1.0$
by increasing the density above the critical filling factor, and conversely
the superfluid order parameter grows from $0.001$ on. For densities
higher than $\rho\approx0.36$ the order parameter of the CDW abruptly vanishes
and the superfluid phase starts to decay.

As the set of variational states is restricted to hard-core bosons
we exclude double occupancy. Hence, for large densities this approach
is not a sufficient choice for calculations of ground states and will not provide
reliable results. Furthermore, the truncation of the Wannier state
-- regarding only leading terms -- also requires adequate low lattice
fillings. Huber et al. \cite{Huber2010} observed a similar behaviour
which might also confirm that we reach a limit of our model at this
density regime and do not observe any physical response.

We tested the stability of the approximation in Eq. \eqref{eq:HeffSmallU}
by taking all coefficients into account up to $0.001U$. We observe
a similar result. The onset of the decline for the superfluid order
parameter remains more or less unchanged with $0.37$. However, the
density for which the order parameter of the CDW will be zero is shifted
to $0.47$. In general, this indicates, nevertheless, the stability
of the truncation procedure.

To conclude, the results for the chequerboard lattice are similar
to the results for the kagom{\'e} lattice derived by Huber et al. \cite{Huber2010},
although, as pointed out above, there are no interstitial sites on
the chequerboard lattice. The essential point is, indeed, that the additional
particles are delocalised. These particles have a high mobility
and this is energetically favourable to localised states, as pointed
out by Huber et al.

\section{Strongly interacting model}
\label{sec:strong-ia}
If $U$ is much larger than $t$, states where a site is occupied
with more than one boson are energetically unfavourable. We take the
limit $U\rightarrow\infty$, i.e. hard-core bosons. For densities
$\rho<1$ this means that we project onto the subspace of multi-particle
states where each site is occupied with at most one boson. Let $P_{\leq1}$
be the projector onto this subspace. Then, the Hamiltonian reads
\begin{equation}
H=tP_{\leq1}\sum_{x,y\in V}a_{xy}b_{x}^{\dagger}b_{y}P_{\leq1}.\label{eq:hard-core-bose_hubbard}
\end{equation}

\subsection{The chequerboard lattice}

The CDW is the exact ground state at the critical density \cite{1112.0131v3}.
Therefore, as in the case of small $U$, we use the CDW at the critical density
and add one further boson. A first variational ansatz for the ground
state is thus 
\begin{align}
|\phi\rangle&= P_{\leq1}\sum_{x\in V}\phi_{x}b_{x}^{\dagger}|F_{1}^{c}\rangle\nonumber \\
&= \sum_{f\in F_{1}^{c}}\frac{1}{\sqrt{|f|}}\sum_{x,x'\in f,x\neq x'}\phi_{x}s_{fx'}b_{x}^{\dagger}b_{x'}^{\dagger}\cdot\nonumber\\
&\quad\cdot\prod_{f'\in F_{1}^{c}\setminus\{f\}}b_{f'}^{\dagger}|0\rangle.\label{eq:cdw+1}
\end{align}
The variation was done on a 20x20 cutout of the chequerboard lattice.
For reasons of numerical stability, we choose a symmetric sub-unit
of 16 lattice sites containing five faces on which $\phi_{x}$ are
varied freely and we let $\phi_{x}=\phi_{R}$ outside that region. The resulting
minimal energy is $-(2|F_{1}^{c}|+0.67)t$ and belongs to a state
where the additional particle is localised on one of the faces of
the sub-unit which are already occupied. It seems energetically favourable to 
disturb only one particle instead of four when putting the additional particle
on a not occupied face. It seems also energetically favourable to have
the additional particle localised and not distributed over many faces.

Since the additional particle is localised on an already
occupied face, a second variational ansatz for the ground state is
\begin{equation}
|u,f\rangle=\sum_{e,e'\in f\in F_{1}^{c}}u_{ee'}b_{e}^{\dagger}b_{e'}^{\dagger}b_{f}|F_{1}^{c}\rangle,
\label{eq:variationalHardCore}
\end{equation}
for an arbitrary face of $F_{1}^{c}$. Since all faces are independent,
this variational problem is equivalent to solving the problem for
two hard-core bosons on a square. The resulting energy is $-2\sqrt{2}t$
for the two hard-core bosons and therefore $-(2|F_{1}^{c}|-2+2\sqrt{2})t=-(2|F_{1}^{c}|+0.83)t$.
If this was the exact ground state, it would be degenerate with a
degeneracy $|F_{1}^{c}|$. As a consequence of this variational ansatz, 
we obtain an exact upper
limit for the ground state energy per particle $e_{0}(\rho)$ for hard-core
bosons on the chequerboard lattice at density $\rho$ to be 
\begin{equation}
e_{0}(\rho)\leq-\big[2\rho_{c}+(2\sqrt{2}-2)(\rho-\rho_{c})\big]t,
\end{equation}
 which holds for $\rho_{c}\leq\rho<2\rho_{c}$. The number of states
at or below that energy must be equal or above the number of occupied
faces.

\begin{figure}[hb]
\begin{centering}
\includegraphics[width=0.3\columnwidth]{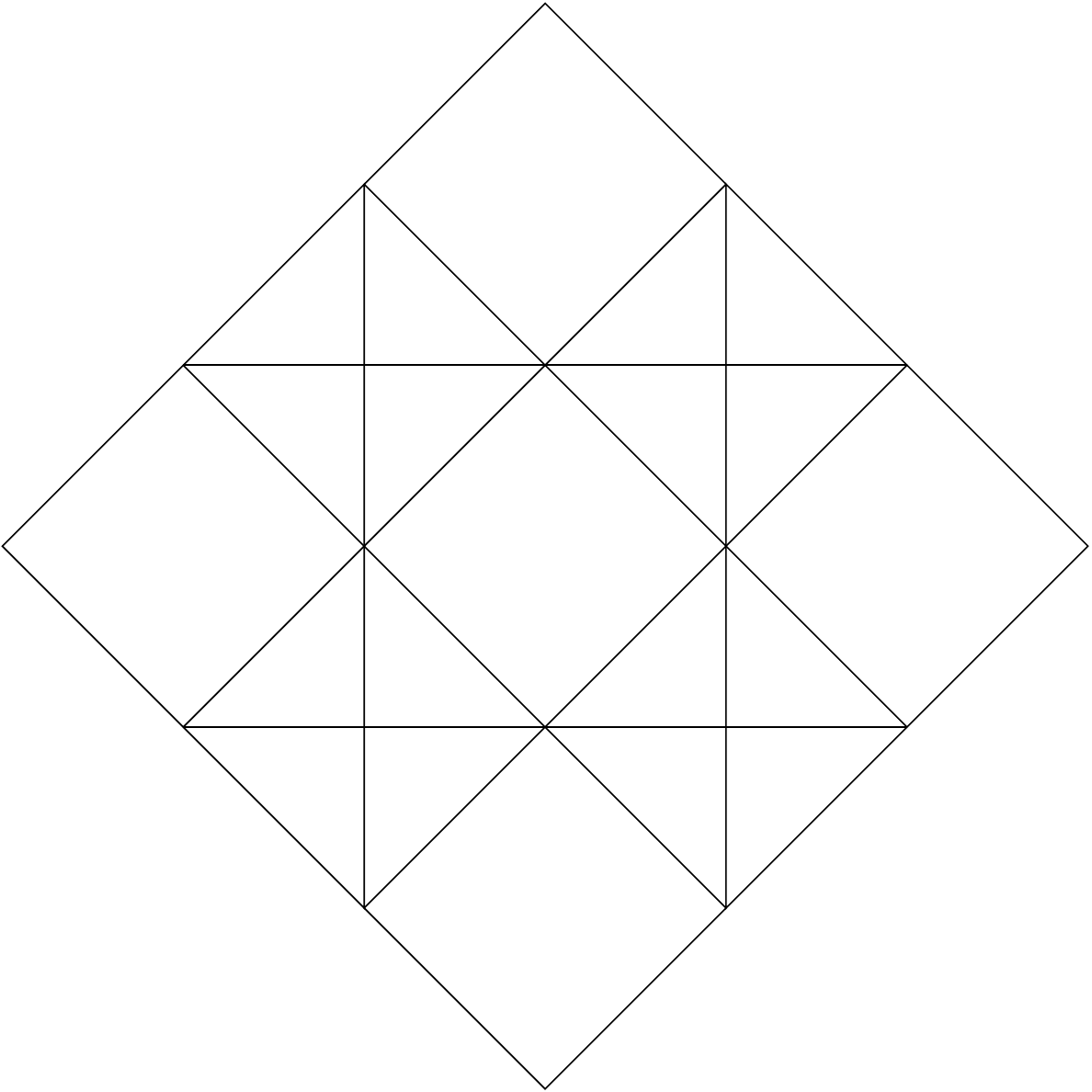}
\par\end{centering}

\protect\caption{\label{fig:Sub-unit-of-the}Sub-unit of the chequerboard lattice. The cycles
denote the occupied faces. In the exact ground state with four particles, each face is 
occupied by one boson. In the variational states (\ref{eq:variationalHardCore}) with 
five bosons one of the faces is occupied by two bosons.}

\end{figure}

As a further test of the ansatz, we look at the next non-trivial sub-unit,
depicted in Fig. \ref{fig:Sub-unit-of-the}. The exact ground state
with four particles is the state where the four faces in the corners
are occupied, depicted as circles in Fig. \ref{fig:Sub-unit-of-the}. 
We now put a fifth boson into the sub-unit. Following
the argument above, we conclude that there are at least four ground states
with an energy below $-(6+2\sqrt{2})t=-8.83t$. A numerical diagonalisation
of the system yields, indeed, exactly four states with energies below
that value. These energies are $-9.016t$, $-9.022t$ (two-fold degenerate),
and $-9.081t$. The corresponding eigenstates have a large overlap
with combinations of the four variational states. They thus show a
similar behaviour as it was observed in different one-dimensional models 
\cite{Takayoshi2013,Tovmasyan2013,Phillips2014}. 
Pairs of bosons are formed.

Since this sub-unit shares no sites with other occupied faces,
we can use this result as a new upper variational limit for densities
between $\rho_{c}$ and $\frac{5}{4}\rho_{c}$, 
\begin{equation}
e_{0}(\rho)\leq-\left[2\rho_{c}+1.081\left(\rho-\rho_{c}\right)\right]t.
\end{equation}
.

\subsection{The kagom{\'e} lattice}

Whereas at weak interactions, physics of hard-core bosons on the chequerboard
and the kagom{\'e} lattice are similar, there may be a difference at strong
interactions. On the kagom{\'e} lattice a CDW ground state at the critical
density has interstitial sites, whereas on the chequerboard lattice
there are no interstitial sites. 

On the kagom{\'e} lattice, the smallest building block is a hexagon. One
particle on the hexagon has an energy of $-2t$. If we put two hard-core
bosons on a hexagon, the energy is $-2\sqrt{3}t$. 

\begin{figure}[hb]
\begin{centering}
\includegraphics[width=0.4\columnwidth]{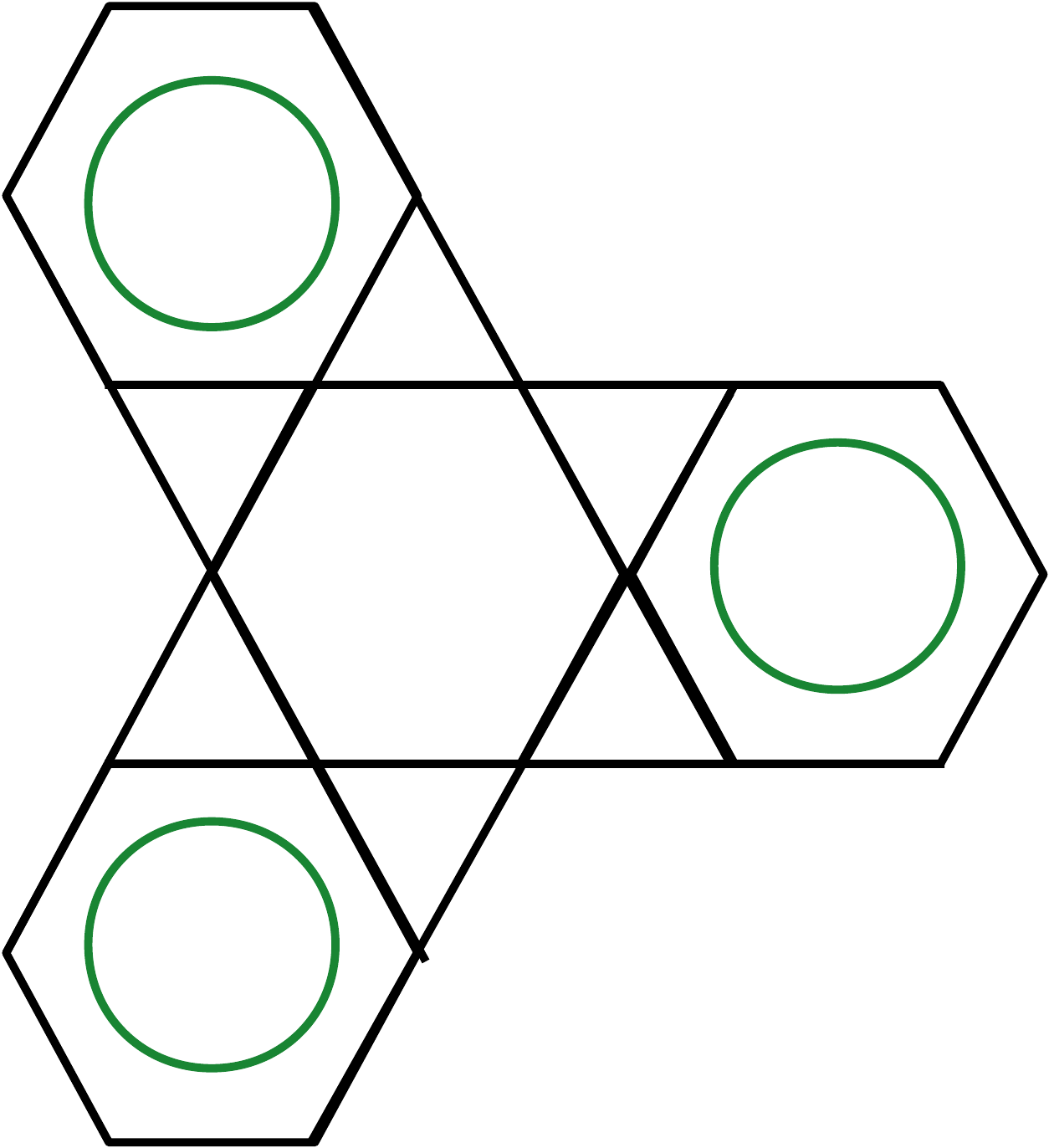}
\par\end{centering}

\protect\caption{\label{fig:fourHexagons}Sub-unit of the kagom{\'e} lattice. The cycles
denote the occupied faces. In the exact ground state with three particles, each face is 
occupied by one boson. In the variational states (\ref{eq:variationalHardCore}) with 
four bosons one of the faces is occupied by two bosons.}

\end{figure}

The smallest non-trivial sub-unit of the kagom{\'e} lattice consists of
one hexagon, sur\-round\-ed by six triangles and three non-touching hexa\-gons,
as depicted in Fig. \ref{fig:fourHexagons}.

Three hard-core bosons on this sub-unit are placed on the three outer
hexagons; the corresponding state is the only ground state on this
sub-unit with energy $-6t$. This state has three interstitial sites.
The ground state with four particles on this sub-unit must have an
energy below $-(4+2\sqrt{3})t=-7.46t$ and following our variational argument, there are at least three
eigenstates below this energy.

A numerical diagonalisation of the sub-unit with four particles shows
that there are three almost degenerate states with energies between
$-7.55t$ and $-7.56t$ and one state, the ground state, with an energy
of $-7.66t$. The ground state has a large expectation value of the
particle number on the three interstitial sites of the inner hexagon,
whereas this expectation value is small on the other three low eigenstates.
The overlap of the ground state, with one of the states where two particles
are on a hexagon, is only 0.085. The other three exact eigenstates
with energies below $-(4+2\sqrt{3})t$ have a high overlap with linear
combinations of the three states with two particles on a hexagon.
The overlap lies above 0.95. This shows, that linear combinations
of the variational states, where the additional hard-core boson is
put on one of the occupied hexagons, describe these three low
energy states well. However, in the true ground state of the sub-unit the
additional particle moves on the interstitial sites and is not localised
on one of the occupied hexagons.

This argument shows that interstitial sites on the kagom{\'e} lattice
are important if hard-core bosons are added to one of the ground states
at the critical filling. This suggests that the argument by Huber
et al. \cite{Huber2010} may remain correct in the regime of strong
interaction. The additional bosons move mainly on the interstitial
sites and form a condensate at densities slightly above the critical
density.

\subsection{Other line graphs}

The same arguments apply to line graphs of other bipartite plane lattices.
If the faces of the underlying bipartite plane lattice form a two-colour
map, the CDW states are at most two-fold degenerate. Since each edge
belongs to two faces, there are no interstitial sites. On the other
hand, if three or more colours are needed to colour the faces of the
underlying bipartite plane lattice, the CDW states have interstitial
sites. This is true for infinite lattices and also for finite sub-units.
In the latter case the outer faces has to be coloured as well. Therefore,
the sub-units should be chosen such that they need the same number
of colours as the original lattice.

An example for the case with more than two colours is the so called
truncated square tiling or $4.8^{2}$-tiling, a bipartite plane lattice
formed of octagons and squares. Hard-core bosons on its line graph
have a unique CDW ground state with one boson on each square. There
are interstitial sites stemming from the additional edges of the octagons.
The smallest non-trivial sub-unit consists of one octagon surrounded
by four non-touching squares. Putting five hard-core bosons on this
sub-unit, we expect at least four ground states with an energy less
than $-(6+2\sqrt{2})t$, corresponding to the four states where the
additional particle is put on one of the already occupied squares.
A numerical diagonalisation of the sub-unit yields eight eigenstates with
an energies less than $-(6+2\sqrt{2})t$. As for the kagom{\'e} lattice,
the ground state has a large expectation value of the particle number
on the four interstitial sites. Let us mention that in this case it
might even be favourable to put two bosons on the interstitial site,
because two hard-core bosons on the octagon have an energy of $-\sqrt{10+2\sqrt{5}}t=-3.804t$. This state contributes to the low lying eigenstates as well. The analysis
on this sub-unit thus shows that the ground state is dominated by
interstitial sites. 

An example where the faces of the underlying bipartite plane lattice
form a two-colour map, like for the square lattice, can be constructed
from the triangular lattice. We put an additional lattice site on
each edge of the triangular lattice. The resulting lattice is a bipartite
plane lattice. Its line graph has hexagons connected via complete
graphs with six vertices, similar to the kagom{\'e} lattice, where the
hexagons are connected via complete graphs with three vertices. Two
particles on a hexagon have the energy $-2\sqrt{3}.$ Let us take
a sub-unit with four hexagons, one in the middle surrounded by three
non-touching hexagons. Since there are three non-touching hexagons,
we expect at least three eigenstates with energies below $-(4+2\sqrt{3})t$.
The exact diagonalisation yields four such eigenstates. The three
lowest states have, again, a large overlap between 0.66 and 0.88 with linear
combinations of the states where the additional particle is put on
a hexagon. The fourth eigenstate appears because of the large boundary
of the sub-unit. Indeed, if we put two particles on the inner hexagon
and two particles on the ring of 12 sites forming the boundary, we
get a variational state with an energy of $-7.36t$, only slightly
above $-(4+2\sqrt{3})t$. This additional variational state has thus
two bound pairs of bosons, in contrast to the three others with one
bound pair, and a large overlap with two of the four low lying eigenstates.
Thus, in the ground state and in the three low lying eigenstates of
this sub-unit, bound pairs of bosons are formed as for the chequerboard
lattice.

There exist one-dimensional systems where these arguments apply as well.
Consider for instance the line graph of the square chain, a lattice
of two chains of corner sharing triangles. The square chain needs
indeed three colours because the two outer faces cannot be coloured
with the same colour as the squares. Thus, the CDW states have interstitial
sites. A numerical diagonalisation shows indeed that for hard-core
bosons, additional particles move on these sites. Let us mention that this
one-dimensional lattice is sometimes called kagom{\'e}-chain, see e.g. 
\cite{Schulenburg2002}

\section{Conclusion and outlook\label{sec:Conclusion-and-outlook}}

We showed that weakly interacting bosons on the chequerboard lattice
behave similarly to weakly interacting bosons on the kagom{\'e} lattice.
We essentially confirmed the results obtained by Huber et al. for
the kagom{\'e} lattice. Above the critical density the additional particles
are delocalised and form a condensate.

For hard-core bosons, the situation is different: our results indicate
that on the chequerboard lattice the additional particles form pairs
of two bound bosons. This is similar to the findings of Phillips et
al. for the saw-tooth chain \cite{Phillips2014}. In contrast to that,
on the kagom{\'e} lattice the additional particles move on interstitial
sites as for weak interaction. 

The kagom{\'e} lattice and the chequerboard lattice are just examples
of two larger classes of two-dimensional lattices with flat bands.
The kagom{\'e} lattice is a prototype of line graphs of bipartite plane
lattices where the faces cannot be coloured with two colours. Instead, three
or four colours are needed. For all lattices in this class, the CDW
states have interstitial sites and we expect that additional particles
will move on these sites, even for strong interactions. The chequerboard
lattice is a prototype of line graphs of bipartite plane lattices
where the faces form a two-colour map. In that case, since each edge
of the original bipartite plane lattice belongs to exactly two faces,
the CDW states have no interstitial sites and pairs of bound particles
are formed for strong interactions.

There are several questions which remain open. For instance, one may
easily construct line graphs of three dimensional lattices like the
line graph of the diamond lattice, an analog to the kagom{\'e} lattice
in three dimensions, or the line graph of the simple cubic lattice.
Partly, such lattices can be found in nature. The line graph of the
diamond lattice is the octahedral sublattice of a spinel. Being line
graphs of bipartite graphs, these lattices have a lowest flat band
and localised single particle eigenstates in that band. As in two
dimensions, it is possible to construct strictly localised states.
As in two dimensions, they are localised on the elementary cycles,
squares or hexagons in the two examples. But unfortunately, a complete
description of all multi-particle ground states at or below the critical
density is not known. The arguments in \cite{1112.0131v3} do not
apply here. Therefore, a discussion of these lattices above the critical
filling is out of reach today. 

Let us mention that hard-core bosons on graphs is a topic that has
recently been discussed in (algebraic) graph theory in connection
with the isomorphism problem \cite{audenaert2005Symsqugra,gamble2010Twoquawalapptograisopro,berry2011TwoquawalEntgraisotes,rudinger2012Nonmulquaranwalapptograisoprostrreggra}.
Spectral properties of hard-core bosons on a graph can reveal certain
properties of the graph which are otherwise difficult to obtain. There
is no direct relation between the current paper and these investigations,
which mainly deal with strongly regular graphs. Nevertheless, it is
interesting to see that here as well a certain spectral property of
hard-core bosons is connected to a graph theoretical problem, here
the colouring of the faces. 

\subsubsection*{Acknowledgement}

A.M. whishes to thank Ehud Altman for stimulating discussions.

Author contribution statement: Both authors contributed equally to the paper.

%
%
%
%

\end{document}